# Observation of pseudogap in $SnSe_2$ atomic layers grown on graphite

Yahui Mao,[1‡] Huan Shan,[1‡] Jinrong Wu,[1] Zejun Li,[1] Changzheng Wu,[1] Xiaofang Zhai,[1,2] Aidi Zhao[1,2*] and Bing Wang[1]

[1] *Hefei National Laboratory for Physical Sciences at the Microscale and Synergetic Innovation Center of Quantum Information and Quantum Physics, University of Science and Technology of China, Hefei, Anhui 230026, China*

[2] *School of Physical Science and Technology, ShanghaiTech University, Shanghai 201210, China;*

*zhaoad@shanghaitech.edu.cn

‡These authors contributed equally to this work

Superconducting metal dichalcogenides (MDCs) present several similarities to the other layered superconductors like cuprates. The superconductivity in atomically thin MDCs has been demonstrated by recent experiments, however, the investigation of the superconductivity intertwined with other orders are scarce. Investigating the pseudogap in atomic layers of MDCs may help to understand the superconducting mechanism for these true two-dimensional (2D) superconducting systems. Herein we report a pseudogap opening in the tunneling spectra of thin layers of $SnSe_2$ epitaxially grown on highly oriented pyrolytic graphite (HOPG) with scanning tunneling microscopy/spectroscopy (STM/STS). A significant V-shaped pseudogap was observed to open near the Fermi level ($E_F$) in the STS. And at elevated temperatures, the gap gradually evolves to a shallow dip. Our experimental observations provide direct evidence of a pseudogap state in the electron-doped $SnSe_2$ atomic layers on the HOPG surface, which may stimulate further exploration of the mechanism of superconductivity at 2D limit in MDCs.

## 1 Introduction

Atomic layers of metal dichalcogenides (MDCs) are receiving intensive attention in recent years due to the intriguing properties dramatically different from their bulk counterparts [1]. Besides single-particle electronic properties, many-body collective phenomena, e.g. charge density waves (CDW) [2–4] and superconductivity [4–10] have been investigated in thin layers of transition metal dichalcogenides (TMDCs) approaching the exact 2D limit. It is worth noting that for high-temperature superconductors, such as superconducting cuprates, in addition to their CDW and superconducting behavior, there are also pseudogaps closely related to superconductivity [11-14]. This phenomenon is similar to some three-dimensional(3D) TMDCs that the existence of CDW, superconductivity and pseudogap can also be observed [15-18]. However, for TMDCs with the thickness of atomic layers, the existence of a pseudogap has only been reported in monolayer $VSe_2$ [19] and single-layer $TiTe_2$ [3]. Comparatively, much less effort has been made in the study of post-transition metal dichalcogenides (PMDCs). It would be highly interesting to explore the possible many-body collective phenomena of PMDCs without the transition metal d-electron contributions to bands near the Fermi surface. Among the PMDCs, $SnSe_2$ is particularly an interesting system because it bears some similarities to the TMDCs indicated from preliminary studies. For bulk $SnSe_2$, with doping or pressurization, the appearance of CDW and superconductivity can be achieved [20-22]. On the other hand, for single-layer $SnSe_2$, the superconductivity with Tc of several Kelvins has also been reported [23, 24]. But whether and how the pseudogap exists in the atomic layers $SnSe_2$ are unknown, which demands immediate experimental investigations.

In this work, we address the above questions by performing the low-temperature STM study on atomically thin $SnSe_2$ layers synthesized on the HOPG surface with van der Waals epitaxy. A striking asymmetric V-shaped gap was observed to open at the $E_F$ for both monolayer (ML) and bilayer (BL) $SnSe_2$, which cannot be accounted for by a single-particle origin. The gap has a magnitude (Δ) of ~ 16meV–22 meV with up to 90% loss of conductance at 5 K and it evolves to a shallow dip at 77 K for the BL $SnSe_2$. The temperature dependence of the nominal gap depth suggests that the gap is not fully closed even at temperature much higher than 77 K. By comparing the tunneling spectra of 2D $SnSe_2$ atomic layers and a bulk $SnSe_2$ sample, we found a considerable amount of charge transferring from the HOPG substrate to $SnSe_2$ layers. It may efficiently lower the conduction band minimum (CBM) of the latter and make the $SnSe_2$ atomic layers conducting. Our experimental study clearly show that a pseudogap exists in the atomically thin $SnSe_2$ layers approaching the 2D limit, which might be closely related to the superconductivity discovered recently in similar systems.

## 2 Materials and Methods

The substrate used for this study was prepared from a piece of freshly cleaved HOPG (Grade SPI-1, SPI Supplies). After the cleavage, the HOPG substrate was immediately transferred into the ultra-high vacuum (UHV) chamber (base pressure $1.5 \times 10^{-10}$ mbar) and degassed at 350 °C for 30 minutes. A clean graphite surface with atomically flat terraces was obtained. $SnSe_2$ films were grown by co-evaporating Sn (99.9999%) and Se (99.999%) with a flux ratio of ~1:50 from an effusion cell (Quantech) and a homemade Knudson cell, respectively. The HOPG substrate was held at room temperature during the deposition. The as-grown $SnSe_2$ film was annealed at 210 °C for 30 minutes in the UHV chamber. The bulk $SnSe_2$ sample was prepared via chemical vapor transport method with iodine as a transport agent. A mixture of Sn, Se and $I_2$ powder was encapsulated in a vacuum quartz tube and placed into a two-zone temperature gradient furnace. The high-temperature zone and low-temperature zone were kept respectively at 800 °C and 700 °C for 1 day, then, both were cooled to room temperature in 2 days. The clean surface of bulk $SnSe_2$ was obtained by cleaving a $SnSe_2$ single crystal with Scotch tape in vacuum (base pressure $1\times10^{-6}$ mbar). All samples of $SnSe_2$ atomic layers on HOPG and bulk $SnSe_2$ were transferred in vacuo into the STM chamber for the low-temperature STM and STS measurements. The STM experiments were carried out with a low-temperature STM (Scienta Omicron GmbH) with a base pressure of $5 \times 10^{-11}$ mbar. A chemically etched tungsten tip was cleaned by $Ar^+$ sputtering prior to all measurements. The $dI/dV$ spectra were recorded with a lock-in amplifier using a sinusoidal modulation of 4 mV and 732 Hz.

## 3 Results and discussion

The schematic structure of the $SnSe_2$ thin film grown on HOPG is shown in Fig. 1a. The $SnSe_2$ layers and underlying HOPG form a typical van der Waals heterostructure. The atomic structure of $SnSe_2$ shares the same arrangement as that of the 1T phase of $TiSe_2$. And each Sn atom is surrounded by two triangles of Se atoms (Fig. 1b), forming a trilayer sandwich structure. Figure 1c is a typical STM topographic image of a $SnSe_2$ film grown on HOPG showing both BL and ML $SnSe_2$. The atomically-resolved STM image taken on the BL $SnSe_2$ (Fig. 1d) shows a hexagonal lattice structure, corresponding to the atomic structure of the top Se layer. It is worth noting that the surface possesses a simple hexagonal lattice and no CDW order can be observed at all sample biases. The lattice constant $a_0$ is experimentally determined to be 3.81 Å for both ML and BL. More structural details of the sample are shown in Supplementary Fig. S1.

The wide-energy-range tunneling spectra taken on the BL $SnSe_2$ show a typical semiconducting behavior (Fig. 1e). The differential conductance is quite low in the bias range of −1.0V to 0V, consistent with the calculated bandgap of the BL $SnSe_2$ [25]. However, when we check carefully the narrow-energy-range spectrum, finite conductance around the $E_F$ and an unexpected gap opening are found. The high-resolution tunneling spectrum near the $E_F$ (Fig. 1f) shows a sharp asymmetric V-shaped gap with two pronounced conductance peaks located at ±16 mV. This energy gap is much larger than the superconducting gaps of the $SnSe_2$ thin layers reported recently [23, 24]. The possible reason for that we didn't observe such a superconducting gap is that the working temperature of our STM is 4.2 K, which is not low enough for observing such a low-temperature superconductivity. This gap feature was robustly observed in all ML and BL terraces with different STM tips, despite the gap behavior being not very uniform across the whole BL surface. The d*I*/d*V* spectra taken on the ML $SnSe_2$ also show a prominent gap opening at $E_F$ but with much stronger spatial inhomogeneity, so here we focus on the gap for BL $SnSe_2$. By analyzing over 300 spectra (taken on BL far away from defects and edges) we found that the magnitude $2\Delta$ varies in a range of ~ 32meV –44 meV and the nominal depth ranges from ~ 65% to 90%, depending on the spatial locations. Fig. 2a shows a set of d*I*/d*V* spectra taken along a 6-nm distance on the BL $SnSe_2$ in a defect-free area. The relative height of the conductance peaks and the symmetry of the gap shape vary from point to point, showing a considerable spatial inhomogeneity and suggesting a short coherence length on the nanometer scale. The short coherence length was also evidenced by the evolution of the gap near the terrace edges. As shown in Fig. 3, two sets of such measurements show that the gap could be locally suppressed near terrace edges, which are strong indications of correlated electronic behavior.

To further verify the origin of the V-shaped gap observed in the BL $SnSe_2$, we studied the temperature dependence of the low-energy tunneling spectra as shown in Fig. 2b. The gap has the largest amplitude at the lowest measured temperature of 5 K and gradually evolves into a slight dip at 77 K. We are not able to derive a temperature dependence of $\Delta$ since the spatial deviation of $\Delta$ is comparable to the temperature-induced variation. However, we found that the depth of the gap shows a monotonic decreasing behavior with increasing temperature. The temperature dependence of the nominal gap depth is plotted in Fig. 2c. The gap depth decreases rapidly from 5 K to 25 K then slowly to 77 K, implying a very high pseudogap temperature $T^*$ (> 77 K), which is much higher than the $T_c$ of superconductivity in $SnSe_2$ bulk or thin layers [20-24].

Such a gap opening in the tunneling spectra of BL $SnSe_2$ strongly suggests a pseudogap nature. First, it has a typical V-shaped behavior which differs from the phonon-induced steplike behavior [26]. Second, the largely dropped conductance in the gap (up to 90% in some areas) suggests that it is unlikely to be a CDW gap which usually accompanies large residual DOS at $E_F$ [18, 27, 28]. This is also consistent with our STM measurements that no CDW order was observed at any bias on the surface. Third, the gap is accompanied with two pronounced side peaks located symmetrically with respect to $E_F$, the characteristic of superconducting or pseudogap coherence peaks [, 29-32], which rule out the possibilities of the Kondo effect. Moreover, the possibility of a superconducting gap is also excluded because the exceptional large gap size will give a huge $T_c$ over 200 K, which cannot be true in this case with residual DOS at $E_F$ at 5 K. All these observations point pseudogap as the most possible origin for the gap.

Many superconductors have a doping phase diagram containing CDW，superconductivity，pseudogap and metal phases, in which the pseudogap region can be reached by tuning the charge-carrier concentration [32-35]. Bulk $SnSe_2$ is an n-type semiconductor with low charge-carrier concentration typically from $10^{17}$ to $10^{18}$ $cm^{-3}$. So there arises a question whether the $SnSe_2$ atomic layers experience a considerable carrier doping on HOPG? We resolve this question by comparing the local work functions between the films and the HOPG substrate first. Supplementary Fig. S2 shows the d*Z*/d*V* spectra taken on HOPG, ML, BL and bulk $SnSe_2$ surfaces. A typical STM image of the bulk sample is shown in Supplementary Fig. S3. The work function differences of the samples can be directly deduced from the differences of the peaks. It is known that sharp resonances, known as the Gundlach oscillations, will appear in the tunneling junction if the applied sample bias voltage is close to or larger than the surface work function. The Gundlach oscillation is a phenomenon of field-emission resonance through standing-wave states in the tip-sample gap, which has been successfully employed to determine the work function differences of thin films. The occurrence of the first resonant peak of Gundlach oscillation corresponds to the onset of the field emission, which reflected the relative amplitude of the work function of the sample. From the d*Z*/d*V* spectra, it is clearly shown that the work function of HOPG ($W_{HOPG}$) is much lower than that of BL ($W_{BL}$) and ML ($W_{ML}$) $SnSe_2$, while bulk $SnSe_2$ has a highest work function ($W_{bulk}$) among all samples. Our results are consistent with earlier reports that bulk $SnSe_2$ has an exceptionally large work function (5.3eV) [36] among MDC semiconductors, while HOPG has a much lower work function of ~ 4.4 eV [37].

Furthermore, we analyzed the CBM of the $SnSe_2$ layers. In Fig. 4a, we show the typical d*I*/d*V* spectra near $E_F$ taken on the surfaces of ML, BL and bulk $SnSe_2$ to compare the thickness dependence of the CBM. All the spectra show pronounced conductance at positive sample biases. The spectrum of the bulk $SnSe_2$ shows nearly zero conductance at negative biases and rise of conductance just above $E_F$, exhibiting the characteristic behavior of heavily doped semiconductors. The spectrum of the BL $SnSe_2$ shows a rise of conductance located at around −60 mV. For ML $SnSe_2$, although a much higher spatial inhomogeneity is found in the tunneling spectra, most spectra show a rise of conductance around −100 mV as demonstrated in the representative spectrum. The CBM of each sample can thus be identified as the onset of the conductance rise in each spectrum, which is largely lowered to below $E_F$ for BL and ML $SnSe_2$, resulting in significantly enhanced DOS at $E_F$. Similarly lowered CBM has been previously observed in superconducting $SnSe_2$–$Co(Cp)_2$ superlattice in which it is 150meV below $E_F$ [20]. The shift of $E_F$ and CBM can be readily understood with an interfacial charge transfer scenario. Fig. 4b shows the schematic band diagrams of a $SnSe_2$ sample before contacting and in contact with a HOPG substrate. As a result of $W_{HOPG} < W_{ML} \sim W_{BL} < W_{bulk}$, the energy of the CBM in $SnSe_2$ is lower than the Fermi energy of HOPG. When $SnSe_2$ layers are in contact with a HOPG, a considerable amount of electrons migrate from the HOPG to the conduction band of $SnSe_2$. This electron transfer lifts the $E_F$ of the interfacial layers of $SnSe_2$ from below CBM to above CBM, turning the layers from semiconducting to metallic. A Schematic diagram of band structures near $E_F$ for bulk,

ML and BL $SnSe_2$ is illustrated in Fig. 4c. The Fermi surface consists of six electron pockets at the M points of the hexagonal Brillouin zone. Such a largely-enhanced DOS at $E_F$ as well as the greatly-increased electron carrier concentration are believed to take responsibility for the emergence of the pseudogap state. To verify the effect of electron doping on the SnSe2 thin layers, we theoretically calculated the density of states (DOS) of a freestanding $SnSe_2$ monolayer without and with electron doping. The calculations do show that the CBM of $SnSe_2$ monolayer can be greatly lowered below $E_F$ even at a small doping level of 0.1 electron per cell, which is fully consistent with the experimental results.

## 4 Conclusion

In summary, we discovered an asymmetric V-shaped pseudogap in the tunneling spectra for $SnSe_2$ atomic layers grown on HOPG. This gap shows similarities to that of pseudogap for some superconductors in shape, energy, spatial inhomogeneity and temperature dependence. We expect more experimental investigations like transport measurements with electrostatic doping [38] and angle-resolved photoemission spectroscopy studies to uncover the nature and physical mechanism behind the pseudogap. The observation and investigation in this work present the first report of pseudogap state in superconducting $SnSe_2$ thin layers, which will greatly help to the future understanding of competing or intertwined orders closely related to superconductivity in MDCs.

## Supplementary Materials

### Supplementary Figures

Fig. S1. Geometric structure of the $SnSe_2$ layers and atomic structure of ML $SnSe_2$.

Fig. S2. $dZ/dV$ spectra taken on HOPG, ML, BL and bulk $SnSe_2$ surfaces.

Fig. S3. Topographic STM image of the surface of a bulk $SnSe_2$ sample.

Fig. S4. Theoretical calculations of DOS of a freestanding $SnSe_2$ monolayer without and with electron doping.

## Acknowledgments

General: We thank Prof. Tao Wu for helpful discussion.

Funding: This work was supported by the National Key R&D Program of China (2016YFA0200603, 2017YFA0205004), the "Strategic Priority Research Program" of CAS (XDB01020100), the National Natural Science Foundation of China (Grants nos.91321309, 21421063, 21473174), and the Fundamental Research Funds for the Central Science Advances Universities (No. WK2060190027, WK2060190084). A.Z acknowledges a fellowship from the Youth Innovation Promotion Association of CAS (2011322).

## References

[1] K. S. Novoselov, A. Mishchenko, A. Carvalho, and A. H. C. Neto, 2D materials and van der Waals heterostructures, *Science* 353(6298), 461 (2016).

[2] P. Chen, Y. H. Chan, X. Y. Fang, Y. Zhang, M. Y. Chou, S. K. Mo, Z. Hussain, A. V. Fedorov, and T. C. Chiang, Charge density wave transition in single-layer titanium diselenide, *Nat. Commun.* 6(1),

1-5 (2015).
[3] P. Chen, W. W. Pai, Y. H. Chan, A. Takayama, C. Z. Xu, A. Karn, S. Hasegawa, M. Y. Chou, S. K. Mo, A. V. Fedorov, and T. C. Chiang, Emergence of charge density waves and a pseudogap in single-layer $TiTe_2$, *Nat. Commun.* 8(1), 1-6 (2017).
[4] M. M. Ugeda, A. J. Bradley, Y. Zhang, S. Onishi, Y. Chen, W. Ruan, C. Ojeda-Aristizabal, H. Ryu, M. T. Edmonds, H. Z. Tsai, A. Riss, S. K. Mo, D. H. Lee, A. Zettl, Z. Hussain, Z. X. Shen, and M. F. Crommie, Characterization of collective ground states in single-layer $NbSe_2$, *Nat. Phys.* 12(1), 92-97 (2016).
[5] X. X. Xi, Z. F. Wang, W. W. Zhao, J. H. Park, K. T. Law, H. Berger, L. Forro, J. Shan, and K. F. Mak, Ising pairing in superconducting $NbSe_2$ atomic layers, *Nat. Phys.* 12(2), 139-143 (2016) .
[6] Y. Cao, A. Mishchenko, G. L. Yu, E. Khestanova, A. P. Rooney, E. Prestat, A. V. Kretinin, P. Blake, M. B. Shalom, C. Woods, J. Chapman, G. Balakrishnan, I. V. Grigorieva, K. S. Novoselov, B. A. Piot, M. Potemski, K. Watanabe, T. Taniguchi, S. J. Haigh, A. K. Geim, and R. V. Gorbachev, Quality heterostructures from two-dimensional crystals unstable in air by their assembly in inert atmosphere, *Nano Lett.* 15(8), 4914-4921 (2015).
[7] E. Navarro-Moratalla, J. O. Island, S. Manas-Valero, E. Pinilla-Cienfuegos, A. Castellanos-Gomez, J. Quereda, G. Rubio-Bollinger, L. Chirolli, J. A. Silva-Guillen, N. Agrait, G. A. Steele, F. Guinea, H. S. van der Zant, and E. Coronado, Enhanced superconductivity in atomically thin $TaS_2$, *Nat. Commun.* 7(1), 1-7 (2016).
[8] J. T. Ye, Y. J. Zhang, R. Akashi, M. S. Bahramy, R. Arita, and Y. Iwasa, Superconducting dome in a gate-tuned band insulator, *Science* 338(6111), 1193-1196 (2012).
[9] J. M. Lu, O. Zheliuk, I. Leermakers, N. F. Q. Yuan, U. Zeitler, K. T. Law, and J. T. Ye, Evidence for two-dimensional Ising superconductivity in gated $MoS_2$, *Science* 350(6266), 1353-1357 (2015).
[10] Y. Saito, Y. Nakamura, M. S. Bahramy, Y. Kohama, J. T. Ye, Y. Kasahara, Y. Nakagawa, M. Onga, M. Tokunaga, T. Nojima, Y. Yanase, and Y. Iwasa, Superconductivity protected by spin-valley locking in ion-gated $MoS_2$, *Nat. Phys.* 12(2), 144-149 (2016).
[11] C. Pepin, V. S. de Carvalho, T. Kloss, and X. Montiel, Pseudogap, charge order, and pairing density wave at the hot spots in cuprate superconductors, *Phys. Rev. B* 90(19), 195207 (2014).
[12] T. Wu, H. Mayaffre, S. Kramer, M. Horvatic, C. Berthier, W. N. Hardy, R. X. Liang, D. A. Bonn, and M. H. Julien, Incipient charge order observed by NMR in the normal state of $YBa_2Cu_3O_y$, *Nat. Commun.* 6, 6438 (2014).
[13] J. J. Wen, H. Huang, S. J. Lee, H. Jang, J. Knight, Y. S. Lee, M. Fujita, K. M. Suzuki, S. Asano, S. A. Kivelson, C. C. Kao, and J. S. Lee, Observation of two types of charge-density-wave orders in superconducting $La_{2-x}Sr_xCuO_4$, *Nat. Commun.* 10, 3269 (2019).
[14] B. Loret, N. Auvray, Y. Gallais, M. Cazayous, A. Forget, D. Colson, M.H. Julien, I. Paul, M. Civelli, and A. Sacuto, Intimate link between charge density wave, pseudogap and superconducting energy scales in cuprates, *Nat. Phys.* 15, 15771-775 (2019).
[15] S. V. Borisenko, A. A. Kordyuk, A. N. Yaresko, V. B. Zabolotnyy, D. S. Inosov, R. Schuster, B. Buchner, R. Weber, R. Follath, L. Patthey, and H. Berger, Pseudogap and charge density waves in two dimensions, *Phys. Rev. Lett.* 100(19), 196402 (2008).
[16] D. V. Evtushinsky, A. A. Kordyuk, V. B. Zabolotnyy, D. S. Inosov, B. Buchner, H. Berger, L. Patthey, R. Follath, and S. V. Borisenko, Pseudogap-driven sign reversal of the Hall effect, *Phys. Rev. Lett*. 100(23), 236402 (2008).
[17] S. V. Borisenko, A. A. Kordyuk, V. B. Zabolotnyy, D. S. Inosov, D. Evtushinsky, B. Buchner, A. N.

Yaresko, A. Varykhalov, R. Follath, W. Eberhardt, L. Patthey, and H. Berger, Two energy gaps and fermi-surface "Arcs" in $NbSe_2$, *Phys. Rev. Lett.* 102(16), 166402 (2009).

[18] A. Soumyanarayanan, M. M. Yee, Y. He, J. van Wezel, D. J. Rahn, K. Rossnagel, E. W. Hudson, M. R. Norman, and J. E. Hoffman, Quantum phase transition from triangular to stripe charge order in $NbSe_2$, *Proc. Natl. Acad. Sci.* USA 110(5), 1623-1627 (2013).

[19] Y. Umemoto, K. Sugawara, Y. Nakata, T. Takahashi, and T. Sato, Pseudogap, Fermi arc, and Peierls-insulating phase induced by 3D-2D crossover in monolayer $VSe_2$, *Nano. Res.* 12(1), 165-169 (2019).

[20] Z. J. Li, Y. C. Zhao, K. J. Mu, H. Shan, Y. Q. Guo, J. J. Wu, Y. Q. Su, Q. R. Wu, Z. Sun, A. D. Zhao, X. F. Cui, C. Z. Wu, and Y. Xie, Molecule-confined engineering toward superconductivity and ferromagnetism in two-dimensional superlattice, *J. Am. Chem. Soc.* 139(45), 16398-16404 (2017).

[21] Y. H. Zhou, B. W. Zhang, X. L. Chen, C. C. Gu, C. An, Y. Zhou, K. M. Cai, Y. F. Yuan, C. H. Chen, H. Wu, R. R. Zhang, C. Y. Park, Y. M. Xiong, X. W. Zhang, K. Y. Wang, and Z. R. Yang, Pressure-induced metallization and robust superconductivity in pristine 1T-$SnSe_2$, *Adv. Electron Mater.* 4(8), 1800155 (2018).

[22] J. W. Zeng, E. Liu, Y. J. Fu, Z. Y. Chen, C. Pan, C. Y. Wang, M. Wang, Y. J. Wang, K. Xu, S. H. Cai, X. X. Yan, Y. Wang, X. W. Liu, P. Wang, S. J. Liang, Y. Cui, H. Y. Hwang, H. T. Yuan, and F. Miao, Gate-induced interfacial superconductivity in 1T-$SnSe_2$, *Nano. Lett.* 18(2), 1410-1415 (2018).

[23] Y. M. Zhang, J. Q. Fan, W. L. Wang, D. Zhang, L. L. Wang, W. Li, K. He, C. L. Song, X. C. Ma, and Q. K. Xue, Observation of interface superconductivity in a $SnSe_2$/epitaxial graphene van der Waals heterostructure, *Phys. Rev. B* 98(22), 220508 (2018).

[24] Z. B. Shao, Z. G. Fu, S. J. Li, Y. Cao, Q. Bian, H. G. Sun, Z. Y. Zhang, H. Gedeon, X. Zhang, L. J. Liu, Z. W. Cheng, F. W. Zheng, P. Zhang, and M. H. Pan, Strongly compressed few-layered $SnSe_2$ films grown on a $SrTiO_3$ substrate: the coexistence of charge ordering and enhanced interfacial superconductivity, *Nano Lett.* 19(8), 5304-5312 (2019).

[25] P. Yu, X. C. Yu, W. L. Lu, H. Lin, L. F. Sun, K. Z. Du, F. C. Liu, W. Fu, Q. S. Zeng, Z. X. Shen, C. H. Jin, Q. J. Wang, and Z. Liu, Fast photoresponse from 1T tin diselenide atomic layers, *Adv. Funct. Mater.* 26(1), 137-145 (2016).

[26] Y. B. Zhang, V. W. Brar, F. Wang, C. Girit, Y. Yayon, M. Panlasigui, A. Zettl, and M. F. Crommie, Giant phonon-induced conductance in scanning tunnelling spectroscopy of gate-tunable graphene, *Nat. Phys.* 4(8), 627-630 (2008).

[27] D. W. Shen, Y. Zhang, L. X. Yang, J. Wei, H. W. Ou, J. K. Dong, B. P. Xie, C. He, J. F. Zhao, B. Zhou, M. Arita, K. Shimada, H. Namatame, M. Taniguchi, J. Shi, and D. L. Feng, Primary role of the barely occupied states in the charge density wave formation of $NbSe_2$, *Phys. Rev. Lett.* 101(22), 226406 (2008).

[28] K. C. Rahnejat, C. A. Howard, N. E. Shuttleworth, S. R. Schofield, K. Iwaya, C. F. Hirjibehedin, C. Renner, G. Aeppli, and M. Ellerby, Charge density waves in the graphene sheets of the superconductor $CaC_6$, *Nat. Commun.* 2(1),1-6 (2011).

[29] Ø. Fischer, M. Kugler, I. Maggio-Aprile, C. Berthod, and C. Renner, Scanning tunneling spectroscopy of high-temperature superconductors, *Rev. Mod. Phys.* 79(1), 353-419 (2007).

[30] T. Kondo, R. Khasanov, T. Takeuchi, J. Schmalian, and A. Kaminski, Competition between the pseudogap and superconductivity in the high-T-c copper oxides, *Nature* 457(7227), 296-300

(2009).

[31] M. J. Lawler, K. Fujita, J. Lee, A. R. Schmidt, Y. Kohsaka, C. K. Kim, H. Eisaki, S. Uchida, J. C. Davis, J. P. Sethna, and E. A. Kim, Intra-unit-cell electronic nematicity of the high-T-c copper-oxide pseudogap states, *Nature* 466(7304), 347-351 (2010).

[32] K. W. Zhang, C. L. Yang, B. Lei, P. C. Lu, X. B. Li, Z. Y. Jia, Y. H. Song, J. Sun, X. H. Chen, J. X. Li, and S. C. Li, Unveiling the charge density wave inhomogeneity and pseudogap state in 1T-$TiSe_2$, *Sci. Bull.* 63(7), 426-432 (2018).

[33] A. Damascelli, Z. Hussain, and Z. X. Shen, Angle-resolved photoemission studies of the cuprate superconductors, *Rev. Mod. Phys.* 75(2), 473-541 (2003).

[334] M. Hashimoto, I. M. Vishik, R. H. He, T. P. Devereaux, and Z. X. Shen, Energy gaps in high-transition-temperature cuprate superconductors, *Nat. Phys.* 10(7), 483-495 (2014).

[35] X. Montiel, T. Kloss, C. Pepin, S. Benhabib, Y. Gallais, and A. Sacuto, η collective mode as A1g Raman resonance in cuprate superconductors, *Phys. Rev. B* 93(2), 024515 (2016).

[36] T. Shimada, F. S. Ohuchi, and B. A. Parkinson, Work function and photothreshold of layered metal dichalcogenides, *Jpn. J. Appl. Phys. 1* 33, 2696-2698 (1994).

[37] H. Ago, T. Kugler, F. Cacialli, K. Petritsch, R. H. Friend, W. R. Salaneck, Y. Ono, T. Yamabe, and K. Tanaka, Work function of purified and oxidised carbon nanotubes, *Synthetic Met.* 103(1-3), 2494-2495 (1999).

[38] Y. Saito, T. Nojima, and Y. Iwasa, Highly crystalline 2D superconductors, *Nat. Rev. Mater.* 2(1), 1-8 (2017).

## Figures

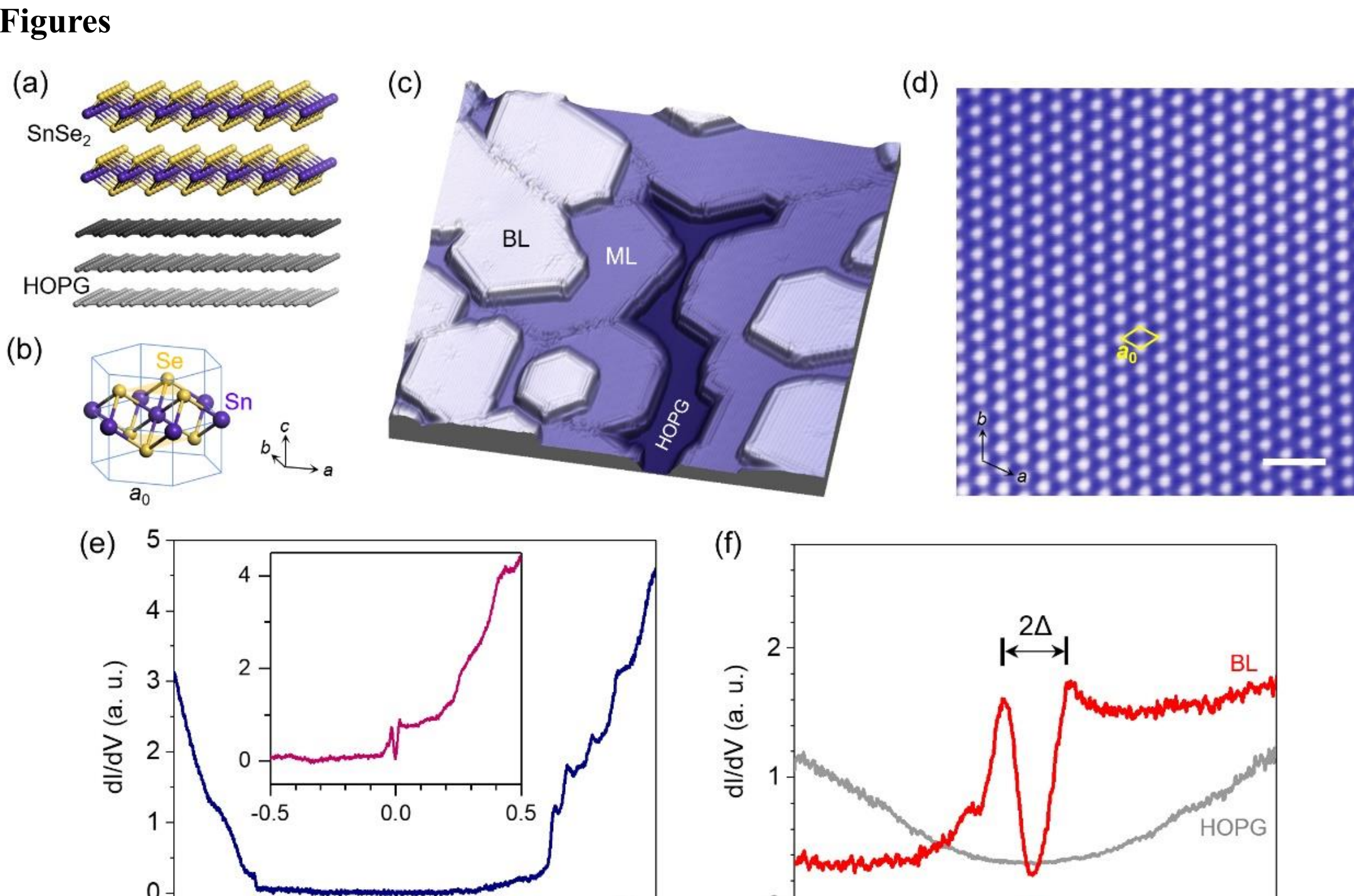

Fig. 1. Structure and STM measurements of $SnSe_2$ atomic layers grown on HOPG. a) Perspective view of the geometric structure of bilayer $SnSe_2$ on HOPG. (b) Crystal structure of a single layer of $SnSe_2$. (c) STM topographic image of a $SnSe_2$ film grown on HOPG showing both BL and ML $SnSe_2$. (55 nm × 55 nm, bias voltage V = 0.5 V, $I_t$ = 100 pA). (d) Atomically-resolved STM image taken on BL $SnSe_2$. The unit cell is denoted by a yellow diamond. (Bias voltage V = 0.5 V, $I_t$ = 50 pA, scale bar: 1 nm.) (e) A typical d*I*/d*V* spectra taken on BL $SnSe_2$. Inset: a narrow-energy-range spectrum showing a gap opening at the $E_F$. The spectrum in the inset was obtained at a much closer setpoints in order to the highlight gap feature. (f) High-resolution tunneling spectrum near the $E_F$, showing the gap opening with 2Δ = 32 meV, where Δ is the energy of the conductance peaks. A tunneling spectrum of HOPG is shown for comparison. All the STM image and spectra were taken at 5 K.

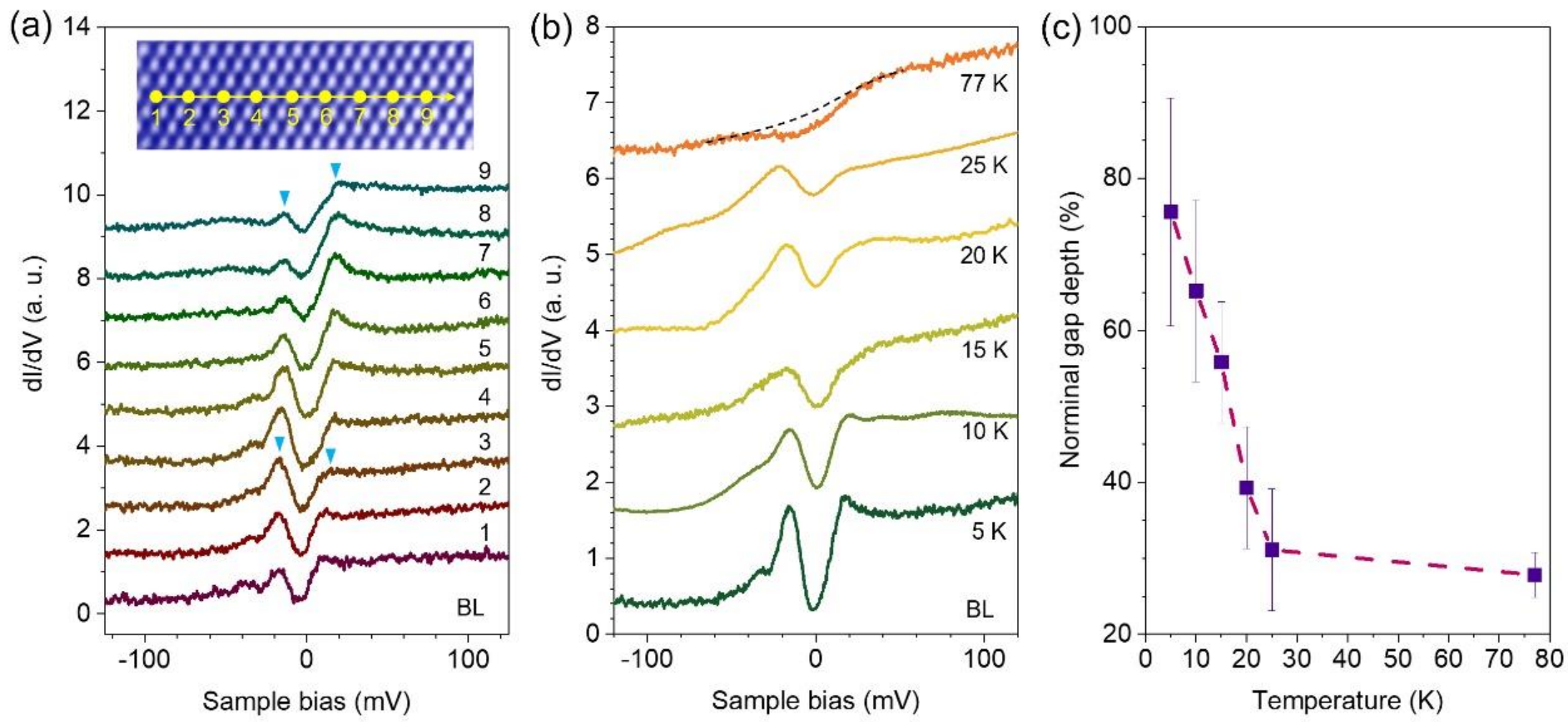

Fig. 2. Spatial variation and temperature dependence of the gap. (a) A set of single-point $dI/dV$ spectra taken on BL along a 6–nm distance at 5 K shows noticeable spatial inhomogeneity of the gap behavior. The blue triangles indicate the change in the relative height of the two side peaks. (Inset: 6.6 nm × 2.2 nm, bias voltage V = 0.5 V, $I_t$ = 300 pA.) (b) Typical spectra taken on BL $SnSe_2$ at different temperatures show the temperature dependence of the asymmetric V-shaped gap. All the spectra are normalized to the difference of conductance at ±100 mV. The depth of the zero-energy gap decreases at elevated temperatures. (c) Nominal gap depth as a function of temperature which decreases monotonically with increasing temperature. The gap depth is defined as 1−ZBC/MCS, where ZBC is the zero-bias conductance, and MCS is the mean conductance of the two shoulders without counting in the contributions from conductance peaks. The error bars represent the min/max values.

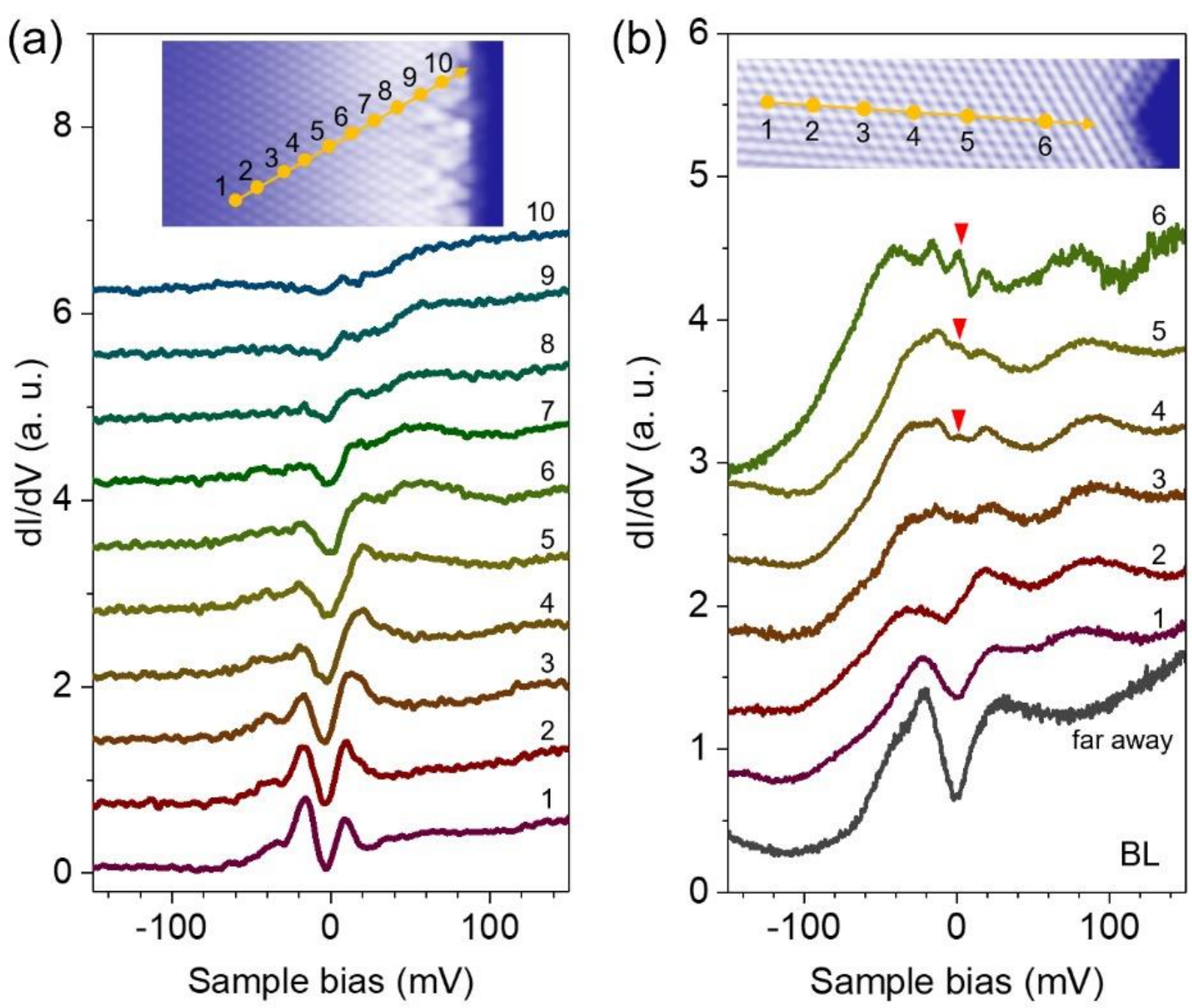

Fig. 3. Tunneling spectra near the terrace edges showing the evolution of the gap. (a) A set of d*I*/d*V* spectra taken on BL $SnSe_2$ close to an edge (inset). The two conductance peaks as well as the gap vanish eventually at the edge. (Inset: 9.0 nm × 5.0 nm, bias voltage V = 0.5 V, $I_t$ = 200 pA.) (b) A set of d*I*/d*V* spectra taken on BL $SnSe_2$ approaching an edge kink (inset), showing the closing of the gap and the arising of a zero-bias conductance peak (red arrows). A spectrum (dark line) taken far away from the edge with a same STM tip is also shown for comparison. (Inset: 11.6 nm × 2.8 nm, bias voltage V = 0.48 V, $I_t$ = 420 pA.) Both sets show strong indications of superconductivity that could be locally suppressed at the edges.

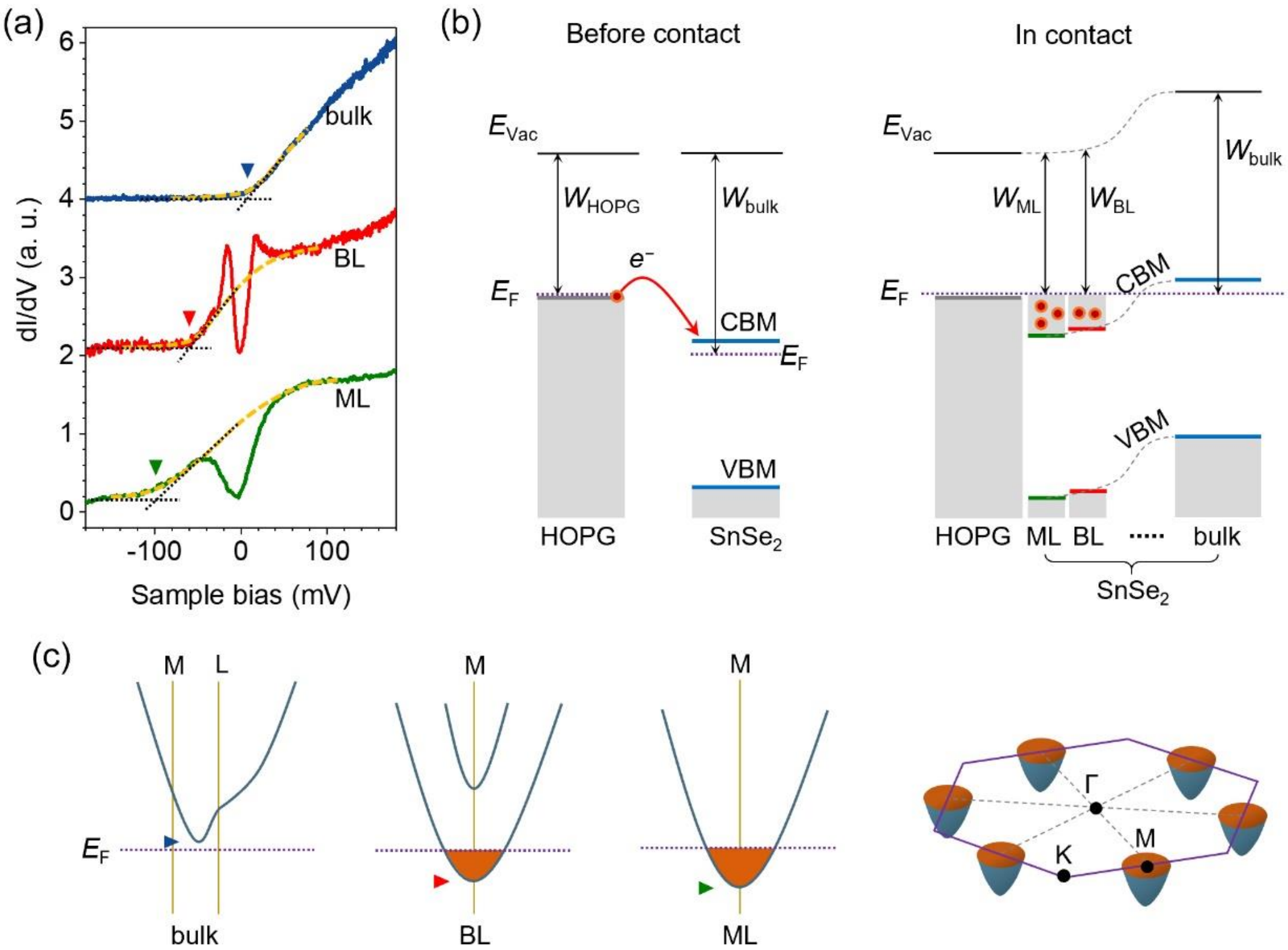

Fig. 4. Thickness dependence of the tunneling spectra and schematic diagrams for the interfacial electron doping. (a) Typical d*I*/d*V* spectra near $E_F$ taken on ML, BL and bulk $SnSe_2$ at 5 K showing the thickness dependence. The CBM is identified as the onset of the conductance rise for all spectra (colored triangles). (b) A schematic energy diagram showing the electrons transferred from HOPG to $SnSe_2$ when $SnSe_2$ is in contact with HOPG. VBM is the valence band maximum. (c) Schematic band structures near $E_F$ for bulk $SnSe_2$ and ML/BL $SnSe_2$ on HOPG. Band structures are derived from recent theoretical calculations [25]. The right panel is a perspective view of the electron pockets at the six M points on the Fermi surface plane of the Brillouin zone for BL and ML.

## Supplementary figures

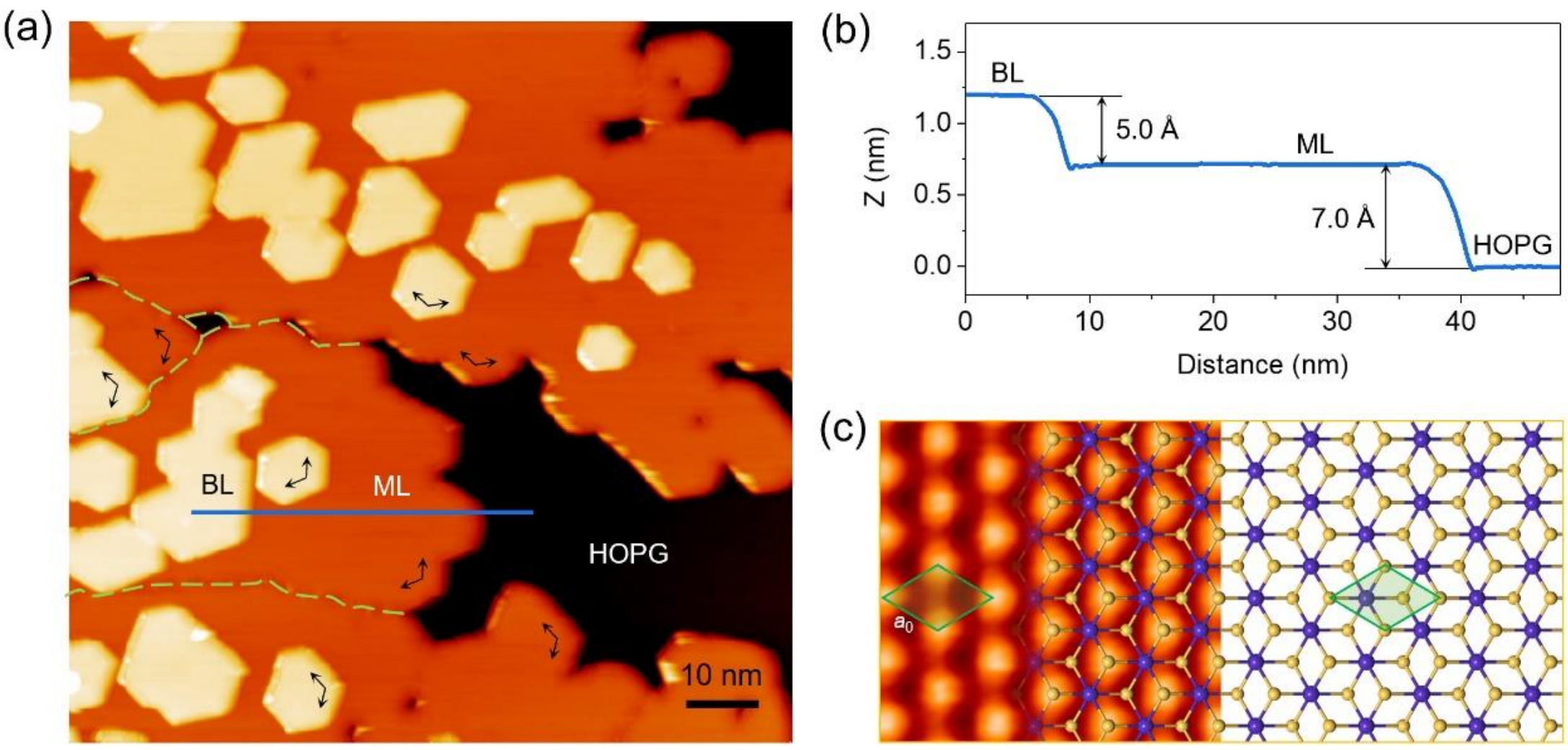

Fig. S1. Geometric structure of the $SnSe_2$ layers and atomic structure of ML $SnSe_2$. (a) The large scale STM image showing four ML $SnSe_2$ domains. The BL $SnSe_2$ possesses the same crystalline orientations with the underlying ML $SnSe_2$ (black arrows), suggesting the optimal stacking structure as shown in Fig. 1A. (Bias voltage V = 1.5 V, $I_t$ = 30 pA) (b) Line profile along the blue line in (a). The apparent heights of ML and BL are 7.0 Å and 5.0 Å, suggesting a weak interaction between ML and HOPG substrate. (c) Atomically-resolved STM image of ML with a crystal structure superimposed on it. The lattice constant of ML is also 3.81 Å, the same with that of BL in Fig. 1d. (STM image: bias voltage V = 0.5 V, $I_t$ = 50 pA.)

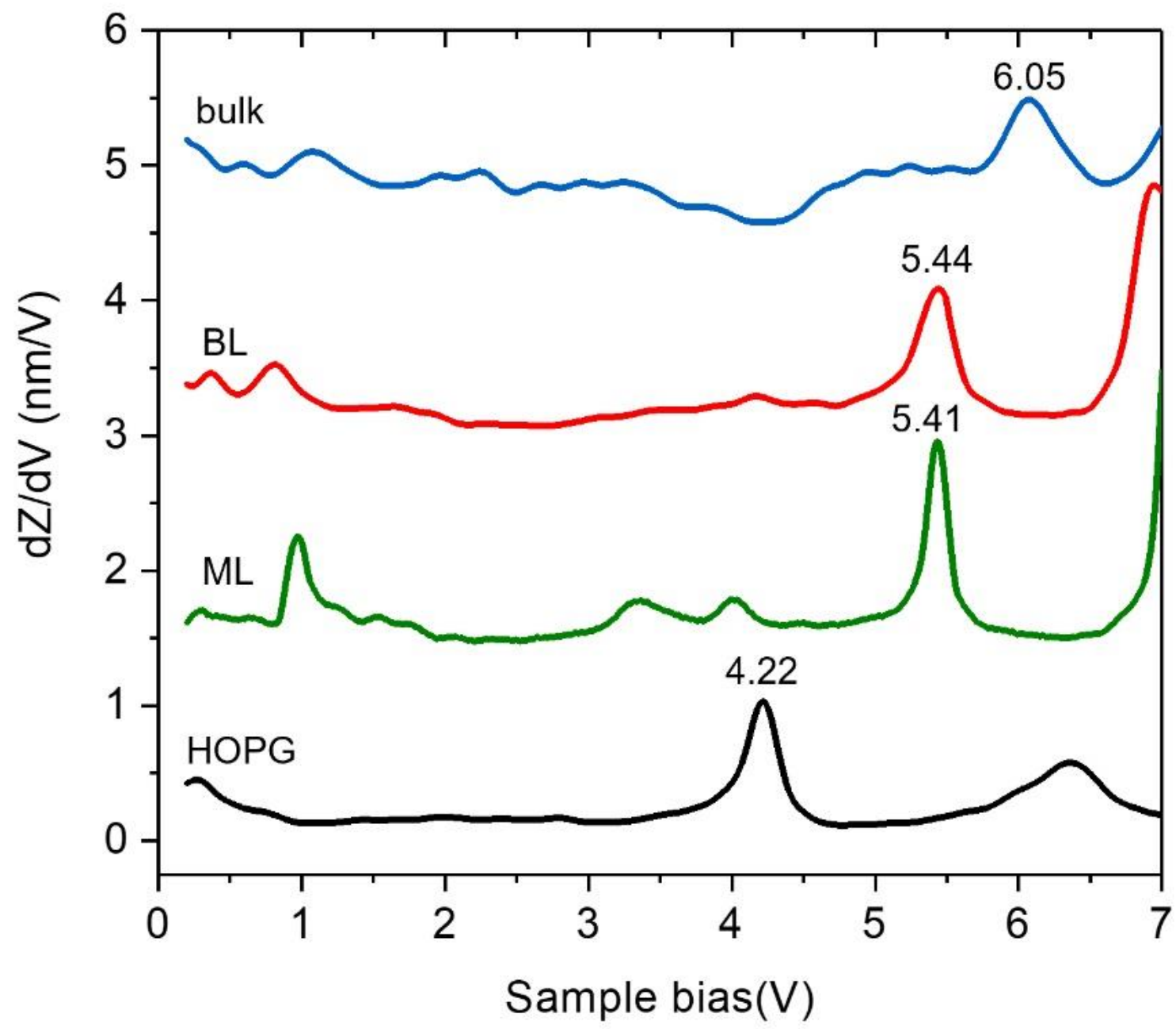

Fig. S2. d$Z$/d$V$ spectra taken on HOPG, ML, BL and bulk $SnSe_2$ surfaces.

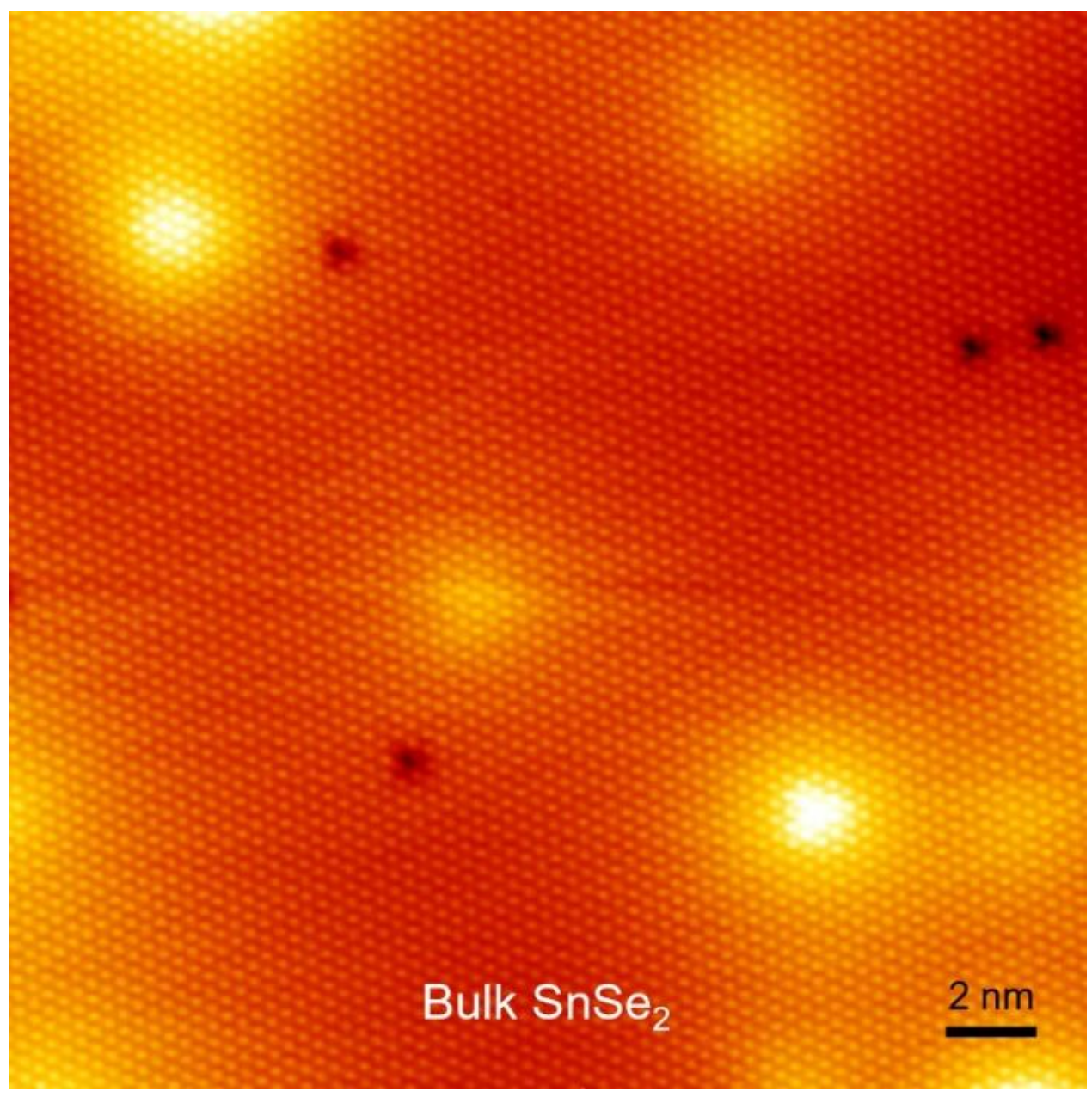

Fig. S3. Topographic STM image of the surface of a bulk $SnSe_2$ sample. The dark spots can be identified as Se vacancies and the bright protrusions are possibly caused by residual iodine intercalated into the layers.

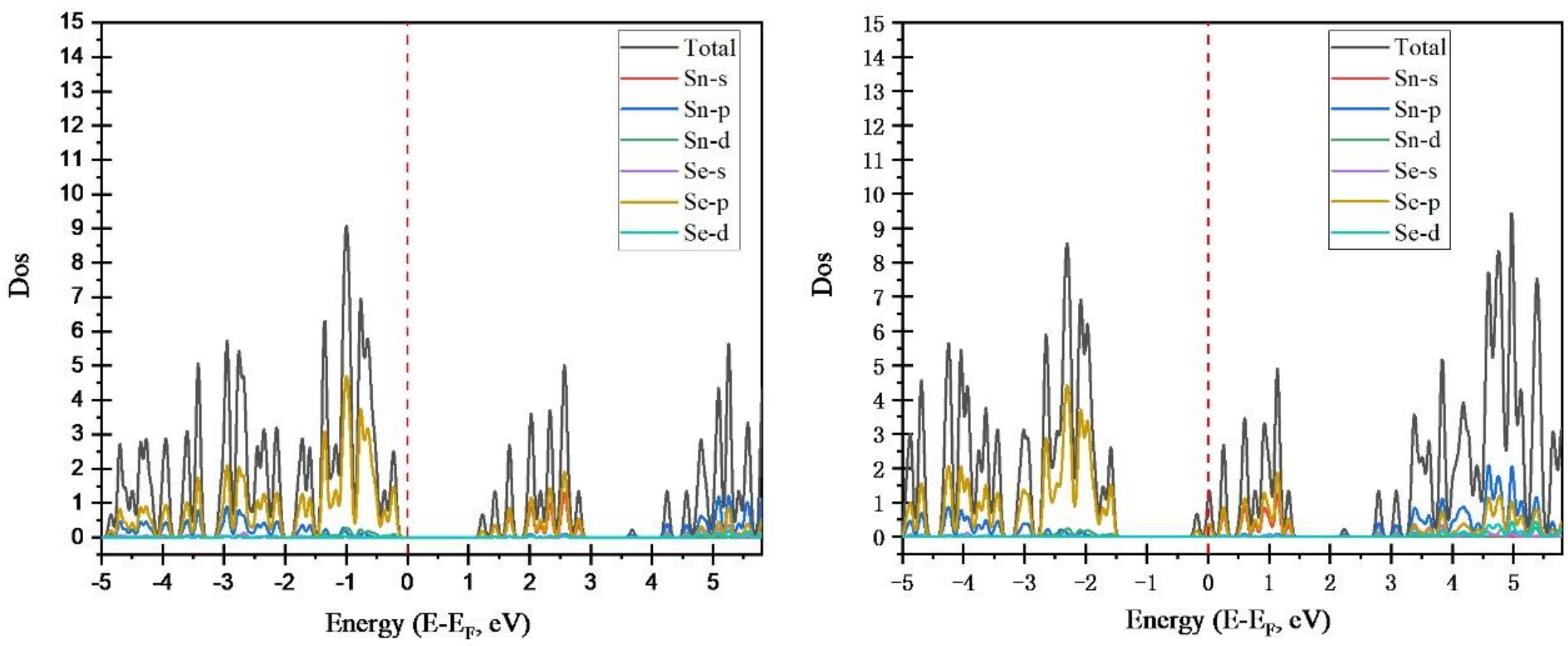

Fig. S4. Density functional theory (DFT) calculations of a freestanding $SnSe_2$ monolayer with HSE06 methods. Left: neutral state; Right: 0.1electron/cell added.